\documentclass[preprintnumbers,nofootinbib,a4paper,10pt]{revtex4-1}

\usepackage[utf8x]{inputenc}

\pdfoutput=1

\usepackage{tabularx} 

\usepackage{tikz}

\usepackage{amsmath,latexsym,amssymb,amsfonts}
\usepackage{float} \usepackage{graphicx}
\usepackage{epstopdf}
\usepackage{graphics}
\usepackage{caption}
\usepackage{subfigure}
\usepackage{bm}
\usepackage{color}

\usepackage{hyperref}
\hypersetup{
	colorlinks=true,
	linkcolor=blue,
	filecolor=magneta,      
	urlcolor=blue,
}

\usepackage[dvips]{psfrag}
\usepackage{ulem}
\usepackage{epsfig}
\usepackage{cases}
\usepackage{tensor}
\usepackage{multirow,array}
\usepackage{booktabs} 

\usepackage{colortbl} 

\newcommand{\cro}{\text{cr}}

\newcommand{\TA}{\text{A}}

\newcommand{\tc}{\tilde{c}}

\newcommand{\TM}{\text{M}}
\newcommand{\TC}{\text{C}}
\newcommand{\TL}{\text{L}}

\newcommand{\GR}{\text{GR}}

\def\lambdabar{\ThisStyle{\ensurestackMath{\stackon[-2.4\LMpt]{%
  \SavedStyle\lambda}{\kern-.5pt\kern\LMpt\rule{1\LMex}{.25pt+.15\LMpt}}}}}

\begin{document}

\title{Cosmic distance duality as a probe of minimally extended varying speed of light}
\author{Seokcheon Lee}
\email[Email:]{skylee2@gmail.com}
\affiliation{Department of Physics, Institute of Basic Science, Sungkyunkwan University, Suwon 16419, Korea}

\begin{abstract}
We obtain the current constraint on the minimally extended varying speed of light (meVSL) model by analyzing cosmic distance duality relation (CCDDR) of it, $D_{L}/D_{A}(1+z)^{-2} = (1+z)^{b/8}$. We use the Pantheon type Ia supernova (SNIa) data, the Hubble parameter $H(z)$ using the cosmic chronometers approach, and the cosmic microwave background (CMB) distance priors from the latest Planck data. We find that the current data show the 1-$\sigma$ deviation from the standard CCDDR. Thus, this provides the constraint on meVSL model and future precision observations might be able to put the stronger constraints on it. 
\end{abstract}

\maketitle


\section{Introduction}
\label{sec:intro}

Etherington shows that a galaxy area distance and an observer area distance are identical up to the redshift factor, $r_{G} = (1+z) r_{Q}$ by assuming that geometric properties are invariant when the roles of the source and observer are transposed \cite{Etherington:1933pm}. This reciprocity theorem is obtained from the integral of the geodesic deviation equation and thus it is valid in any space-time as long as photons follow null geodesic and the geodesic deviation equation holds. If photon numbers are conserved, then one can obtain the cosmic distance duality relation (CCDDR) from the reciprocity theorem by relating these area distances to angular and luminosity distances \cite{Ellis:1998ct,Ellis:2007grg}. Thus, CCDDR might be used to test the validity of the current standard model of cosmology, called the $\Lambda$CDM model irrelevant to its space-time background. 

Both the angular diameter distance and the luminosity distance are complimentarily used to obtain values of cosmological parameters in modern cosmology. This analysis relies on the validity of CCDDR and thus it is quite important to investigate its validity. There are various possibilities for the violation of one or more of these conditions. The several known mechanisms for this are the following. 

 \begin{itemize}
 	\item Violation of photon number conservation: photon absorption by dust, photon decay, or photon mixing with other light particles reduce the number of photons \cite{Csaki:2001yk,Bassett:2003zw,Lima:2011ye}.
	\item Deviation from a metric theory of gravity: non-minimal coupling gravity induces photons not to follow null geodesic \cite{Santana:2017zvy,Belgacem:2017ihm,Azevedo:2021npm}.
	\item Photons do not travel on null geodesics due to kinematics: torsion or birefringence \cite{Schuller:2017dfj,Schneider:2017tuk}. 
	\item Gravitational lensing  (GL): GL causes a dispersion in Hubble-diagram when measured by the finite-sized light bundle \cite{Amanullah:2002xh}.   
	\item Minimally extended varying speed of light (meVSL): due to the modification of the photon frequency in this model, it induces the modification on the CCDDR as $D_{L}/D_{A} = (1+z)^{2-b/8}$ \cite{Lee:2020zts}.
 \end{itemize}

Even though the standard model of cosmology has passed the enormous observations, there are still open problems related to its theoretical and experimental aspects ( for recent reviews, see Ref. \cite{Bull:2015stt,Perivolaropoulos:2021jda}). Thus, one needs to keep investigating alternatives for $\Lambda$CDM. Recently, we have proposed one of cosmologically varying speed of light models as a minimal extension of the standard cosmology \cite{Lee:2020zts}. It preserves the local Lorentz invariant with accompanying variations of other physical constants related to the whole set of physical equations. 
  
The CCDDR is given by 
\begin{align}
\frac{d_{L}}{d_{A}} (1 + z)^{-2} = 1 \label{cCDDR} \,,
\end{align}
where $d_{L}$ and $d_{A}$ are the luminosity distance and the angular diameter distance, respectively. Thus, this relationship can be in principle testable by measuring the source whose intrinsic luminosities (standard candles) as well as their intrinsic sizes (standard rulers) are known. Even though ideally both quantities must be measured in a model-independent way, one must still rely on cosmological observational values based on a specific cosmological model to testify the validity of CCDDR due to the limitations in each measurement. There have been several tests for CCDDR by using various astrophysical/cosmological observations. 

 Any deviation of CMB spectrum from Planck spectrum occurs the violation of CCDDR \cite{Ellis:2013cu}. Gamma-ray bursts (GRBs) can be used as standard candles and provide luminosity distances at high redshifts. They have been used to test the validity of CCDDR \cite{Holanda:2014lna}.  One can measure the luminosity standard siren from gravitational waves (GWs) without the need for a cosmic distance ladder \cite{Yang:2017bkv,Qi:2019spg,Arjona:2020axn}. This provides a unique CCDDR test. The type Ia supernovae (SNIa) data also have been used as a luminosity distance to test CCDDR \cite{Bassett:2003vu}. Interestingly, strongly lensed SNIa can provide both the luminosity and the angular diameter distance simultaneously to provide a useful CCDDR test \cite{Renzi:2020bvl}. If one uses the 21cm radio of HI emission line, then one can ignore the dimming of the light source due to dust. Thus, a 21cm hyperfine transition signal can offer another unique CCDDR test
\cite{Khedekar:2011gf}. From galaxy clusters (GCs), one can measure the angular diameter distance (ADD) by using the Sunyaev-Zeldovich effect (SZE) and X-ray surface brightness. This also yields a validity test for CCDDR \cite{Uzan:2004my,DeBernardis:2006ii}. One can also use the fact that the relation between the gas mass fraction of GCs from SZE ($f_{\text{SZE}}$) and that from X-ray surface brightness ($f_{\text{X}}$)  is given by $f_{\text{SZE}}/f_{\text{X}} = d_{L}/(d_{A} (1+z)^2)$ \cite{Holanda:2012at}. One can extend this method by adopting cosmological parameters from cosmic microwave background (CMB) anisotropy observation in addition to the morphology of GCs  \cite{Holanda:2010ay}. The model-independent test by using measurements of the Hubble parameter $H(z)$ along with GCs also provides an alternative test for CCDDR \cite{Santos-da-Costa:2015kmv}. Another analysis has been done by using the luminosity distance and angular diameter distance from SNIa and GCs, respectively \cite{DeBernardis:2006ii,Holanda:2010vb,Li:2011exa,Meng:2012apj,Holanda:2012aa,Goncalves:2011ha,Yang:2013coa}. Angular diameter distance obtained from baryon acoustic oscillation (BAO) combined with the luminosity distance from SNIa has also been investigated for the validity check for CCDDR \cite{More:2008uq,Nair:2012dc,Wu:2015prd,Ma:2016bjt,Martinelli:2020hud}. Both angular diameter distance and luminosity distance are functions of the Hubble parameter and the growing amount of its measurements on various redshift can be also complementarily used in order to test the validity of CCDDR \cite{Avgoustidis:2009ai,Avgoustidis:2010ju,Holanda:2012ia}. A method to test CCDDR based on strong gravitational lensing (SGL) provides free from all other prior assumptions related to details of the cosmological model \cite{Liao:2015uzb,Holanda:2015zpz}. Techniques obtaining the angular diameter distance from the SGL and the luminosity distance from GWs are also introduced to test the validity of CCDDR \cite{Liao:2019xug}. Because both GWs and SNIa can measure the luminosity distance in similar redshift ranges and thus provide a method to constrain the cosmic opacity \cite{Wei:2019fwp}. There have been several validity tests of CCDDR by using combinations of different observations like combined observations of SGL and Hubble parameter constructed from HII galaxy \cite{Ruan:2018dls}, the combination of BAO, CMB, and SNIa \cite{Lazkoz:2007cc},  merging of BAO, GWs, and SNIa \cite{Hogg:2020ktc}, the amalgamation of GCs, H, and SNIa \cite{Bora:2021cjl}, etc.  

The outline of this manuscript is as follows. In Section \ref{sec:Rev}, we briefly review equations for various geometrical distances of meVSL model. We also compare the CCDDR relation in meVSL with various parameterizations of it used in literature. We analyze data to obtain cosmological parameters and time evolutions of the speed of light and the gravitational constant in Section \ref{sec:Anal}. In Section \ref{sec:Con}, we conclude and summarize. 

\section{Review}
\label{sec:Rev}

We briefly review various cosmological distances defined in the meVSL model \cite{Lee:2020zts}. We limit our consideration to the flat universe in this manuscript. 

\subsection{Distances in meVSL}
\label{subsec:meVSL}

Both the Hubble and the acceleration of expansion of meVSL model are obtained from Friedmann equations
\begin{align}
H^2 &= \left( \frac{\dot{a}}{a} \right)^2 = \frac{8 \pi G_0}{3} \left( \rho_{m0} a^{-3} + \rho_{X0} a^{-3(1+\omega)} \right) a^{b/2} \label{H2} \,, \\
E^2 &= H^2/H_0^2 = \left( \Omega_{m0} a^{-3} + \Omega_{X0} a^{-3(1+\omega)} \right) a^{b/2} \equiv E^{(GR)^2} a^{b/2} \label{E2} \,, \\
\frac{\ddot{a}}{a} &= - \frac{4 \pi G_0}{3} \left[ \left( 1 - \frac{b}{2} \right) \rho_{m0} a^{-3} + \left( 1 + 3 \omega - \frac{b}{2} \right) \rho_{X0} a^{-3(1+\omega)} \right] a^{b/2} \label{ddota} \,, 
\end{align}
where a subscript 0 on each parameter denotes its value at the present epoch, $G_0$ is the present value of the gravitational constant, $\rho_{m 0}$ and $\rho_{X 0}$ are present values of mass densities of matter and dark energy, $H_0$ is the present value of the Hubble parameter, $\Omega_{i0} \equiv \rho_{i 0} / \rho_{\cro 0}$ is the present mass density contrast of i-component, and $\omega_i$ is its equation of state (e.o.s), and $b$ is a model parameter of meVSL given by $\tilde{c} (z) = \tilde{c}_0 (1+z)^{-b/4}$. The comoving distance $D_{\TC}$, the transverse comoving distance $D_{\TM}$, and the angular diameter distance $D_{\TA}$ in meVSL model are given by \cite{Lee:2020zts}
\begin{align}
 D_{\text{C}}(z) &\equiv \int_{0}^{r} \frac{dr'}{\sqrt{1-kr^2}} = \frac{\tc_0}{H_0} \int_{0}^{z} \frac{dz'}{E^{(\GR)}(z')}  \quad \text{where} \quad  E^{2} \equiv \left( \sum_{i} \Omega_{i0} a^{-3(1+\omega_i)} \right)  \label{Dcmp} \,, \\
 D_{\text{M}}(z) &=  D_{\text{C}}(z) \quad , \quad D_{\TA}(z)  = (1+z)^{-1} D_{\TM}(z) \label{DA} \,,
 \end{align}	
 where $\tc_0$ is the present value of the speed of light. Thus, both the comoving distance and the angular diameter distances of meVSL model are equal to those of GR. However, the relation between the luminosity distance and the transverse comoving distance is modified due to the modification of the photon frequency as
\begin{align}
D_{\TL}(z) = \left( 1 + z \right)^{1 - \frac{b}{8}} D_{\TM}(z) = \left( 1 + z \right)^{2 - \frac{b}{8}} D_{\TA}(z) \label{CCDDRmeVSL} \,.
\end{align}
Thus, the CCDDR of the meVSL model is written as 
\begin{align}
\eta(z) \equiv \frac{D_{\TL}}{D_{\TA} \left( 1 + z \right)^2} = \left( 1 + z \right)^{-\frac{b}{8}} \label{etameVSL} \,.
\end{align}
 
\subsection{Parametrizations of CCDDR deviation}
\label{subsec:etaz}

 After Gordon introduces the concept of an optical metric by using a refraction index \cite{Gordon:1923AdP}, Chen and Kantowski generalize it by including an absorption phenomenon \cite{Chen:2009prd1, Chen:2009prd2}. This light absorption violates the photon number conservation law and thus modifies the luminosity distance 
 \begin{align}
 D_{\TL}(z) \equiv e^{\frac{\tau}{2}} D_{\TL}^{(\GR)}(z) = e^{\frac{\tau}{2}} (1+z) D_{\text{C}}(z) \label{DLChen} \,,
 \end{align}
 where $\tau$ represents the optical depth related to the cosmic absorption of the photon. This can be rewritten by introducing the dimensionless cosmic absorption parameter, $\alpha_{\ast}$ as \cite{Chen:2009prd1, Chen:2009prd2}
 \begin{align}
 \tau(z) \equiv \int_{0}^{z} \frac{\alpha_{\ast}}{(1+z') E(z')} dz' \label{tauchen} \,.
 \end{align}
In this derivation, one adopts the assumption of no cosmic refraction and thus the angular diameter distance is the same as that of GR. Thus, this induces the deformation of CCDDR which can be represented by the parameterization of the deformation function, $\eta(z) = e^{\tau(z)/2}$. There have been several one or two index parameterizations of $\eta(z)$, $\eta_0 + \eta_i z$, $\eta_0 + \eta_i z/(1+z)$, $\eta_0 + \eta_i z/(1+z)^2$, $\eta_0 - \eta_i \ln(1+z)$, $\eta_0/(1+z)$, $\eta_0/(1+z) \exp[z/(1+z)]$, $(1+z)^{\epsilon}$, etc \cite{Meng:2012apj,Nair:2011dp}. 

We summarize the current constraints on a possible departure from CCDDR by using various parameterizations from different observables in Table~\ref{tab:my-table1}.

\begin{table}[]
\caption{A summary of the current constraints on CDDR parameters from various observations. ADD denotes angular diameter distance obtained from X-ray and SZE. BAOcom includes 6dFGS \cite{Beutler:2011mnras}, SDSS DR7 LRG BOSS CMASS \cite{Xu:2012mnras}, WiggleZ \cite{Blake:2012mnras}, SDSS DR10, 11 \cite{BOSS:2013rlg}, SDSS DR7 MGS \cite{Ross:2014qpa}, SDSSIII DR12 BOSS \cite{Gil-Marin:2015nqa}, DR14 QSOs \cite{Ata:2017dya}, DES \cite{DES:2017rfo}, DR14 LRG \cite{Bautista:2017wwp}, and DR14 QSOs \cite{Blomqvist:2019rah}. $H$com is composed of LRGs \cite{Gaztanaga:2008xz}, Cosmic Chronometers (CC) \cite{Stern:2010apjs}, CC \cite{Moresco:2012jcap}, WiggleZ \cite{Blake:2012mnras}, and LRGs \cite{Zhang:2014aa}. The asterisk symbol means the 2-$\sigma$ error bars.}
\label{tab:my-table1}
\scriptsize 
\begin{tabular}{|c|c|c|c|c|c|c|}
\hline
  Probes   & Data Sample    &  $1+ \eta_0 z$ &  $1+ \eta_0 \frac{z}{(1+z)}$     & $(1+z)^{2+\epsilon}$     &  $\eta_0$    &   Ref                     \\ \hline
 21cm      &                         &&                      &            $\Delta \epsilon$ = 0.010             &          &         \cite{Khedekar:2011gf}        \\ \hline
 \multirow{4}{*}{GCs}      &      ADD \cite{Reese:2002sh}                          & &               &                         &       $0.87^{+0.04}_{-0.03}$ (18)   &  \cite{Uzan:2004my}          \\ 
&      ADD \cite{Bonamente:2005ct}                           & &               &                         &        $0.97^{+0.03}_{-0.03}$ (38)   &  \cite{DeBernardis:2006ii}          \\ 
       &    \multirow{2}{*}{gas mass fraction \cite{LaRoque:2006te} $^{\ast}$}                                     & $-0.15 \pm 0.14$ (38) & $-0.22 \pm 0.21$         &                         &                    &  \multirow{2}{*}{\cite{Holanda:2012at} }    \\ 
              &                                   & $-0.06 \pm 0.16$ (29) & $-0.07 \pm 0.24$         &                         &                    &       \\ \hline
\multirow{2}{*}{GCs + CMB} & \multirow{2}{*}{ADD \cite{Bonamente:2005ct,DeFilippis:2005hx} + WMAP7 \cite{Komatsu:2010apjs}} & $-0.056 \pm 0.1$ (el)& $-0.088 \pm 0.14$ && & \multirow{2}{*}{\cite{Holanda:2010ay}} \\ 
&& $-0.12 \pm 0.055$ (sp) & $-0.175 \pm 0.083$  && & \\ \hline
\multirow{2}{*}{GCs + H} & ADD \cite{Bonamente:2005ct} + Hcom & $-0.100^{+0.117}_{-0.126}$ & $-0.157^{+0.179}_{-0.192}$ && & \multirow{2}{*}{\cite{Santos-da-Costa:2015kmv}} \\ 
& gass mass fraction \cite{LaRoque:2006te} + Hcom & $0.062^{+0.168}_{-0.146}$ & $-0.166^{+0.337}_{-0.278}$  && & \\ \hline
\multirow{9}{*}{GCs + SNIa} & ADD \cite{Bonamente:2005ct} + Gold \cite{SupernovaSearchTeam:2004lze} + SNLS  \cite{SNLS:2005qlf}  & &&& $ 1.01 \pm 0.07 $   & \cite{DeBernardis:2006ii} \\ 
&  ADD \cite{Bonamente:2005ct,DeFilippis:2005hx} + Constitution \cite{Hicken:2009apj}$^{\ast}$ & $-0.28 \pm 0.44$ &$-0.43 \pm 0.60$ & &  & \cite{Holanda:2010vb} \\
&  ADD \cite{Bonamente:2005ct,DeFilippis:2005hx}  + Constitution \cite{Hicken:2009apj} + Union2 \cite{Amanullah:2010apj}$^{\ast}$& $-0.37 \pm 0.35$ & $ -0.56 \pm 0.49$  && & \cite{Li:2011exa} \\ 
& \multirow{2}{*}{ADD \cite{Bonamente:2005ct,DeFilippis:2005hx}  +  Union2 \cite{Amanullah:2010apj}} & $0.021 \pm 0.179$ (el) & $-0.028 \pm 0.214$ && & \multirow{2}{*}{\cite{Meng:2012apj}}\\ 
& & $-0.287 \pm 0.136$ (sp) & $0.337 \pm 0.163$ && & \\ 
& \multirow{2}{*}{ADD \cite{LaRoque:2006te,Ettori:2009aa} + Union2 \cite{Amanullah:2010apj}$^{\ast}$} & $-0.03^{+1.03}_{-0.65}$ (La) & $-0.08^{+2.28}_{-1.22}$ && & \multirow{2}{*}{\cite{Goncalves:2011ha}} \\ 
& & $-0.97^{+0.54}_{-0.38}$ (Et) & $-1.60^{+0.90}_{-0.70}$ && & \\  
&  \multirow{2}{*}{ADD \cite{Bonamente:2005ct,DeFilippis:2005hx}  + Union2 \cite{Amanullah:2010apj}} & $0.16^{+0.56}_{-0.39}$ (el) &  && & \multirow{2}{*}{\cite{Yang:2013coa}} \\  
& & $0.12^{+0.20}_{-0.17}$ (sp) & && & \\ \hline
\multirow{6}{*}{SNIa + BAO} & Union2 \cite{Amanullah:2010apj} + SDSS \cite{Percival:2009mnras} + 6dFGS \cite{Beutler:2011mnras} & \multirow{2}{*}{$-0.066 \pm 0.070$} & \multirow{2}{*}{$-0.103 \pm 0.106$} && &\multirow{2}{*}{\cite{Nair:2012dc} }\\
& + WiggleZ \cite{Blake:2011mnras2} + BOSS \cite{Anderson:2012mnras}  && && & \\
 & Union2.1 \cite{Suzuki:2012apj} + WiggleZ \cite{Blake:2012mnras}  & $-0.086 \pm 0.064$ (h=0.700) & $-0.131\pm0.098$ && &\multirow{2}{*}{\cite{Wu:2015prd}} \\
  &+ DR7 \cite{Xu:2013mnras} + DR11 \cite{Samushia:2013yga}  & $-0.027\pm0.064$ (h=0.678) & $-0.039 \pm 0.099$ && &\\
&  JLA \cite{SDSS:2014iwm} + WiggleZ \cite{Blake:2011mnras,Blake:2012mnras} + BOSS \cite{BOSS:2016wmc}                                           &             &&         & $1.02 \pm 0.08$  &   \cite{Ma:2016bjt}     \\
& Pantheon \cite{Scolnic:2017caz} + BAOcom && & $0.013 \pm 0.029$ & & \cite{Martinelli:2020hud}\\  \hline
\multirow{3}{*}{SNIa + H}   & Union \cite{SupernovaCosmologyProject:2008ojh} + LRGs \cite{Jimenez:2003iv,Simon:2004tf}$^{\ast}$  && & $-0.01^{+0.08}_{-0.09}$  & & \cite{Avgoustidis:2009ai} \\ 
&  Union \cite{SupernovaCosmologyProject:2008ojh} + HST \cite{Riess:2009apj} + Keck \cite{Stern:2009jcap}$^{\ast}$ && & $-0.04^{+0.08}_{-0.07}$  & &  \cite{Avgoustidis:2010ju} \\
& Union2.1 \cite{Suzuki:2012apj} + SCP-0401 \cite{Rubin:2013apj} + H(z)com & & & $0.017 \pm 0.055$ & & \cite{Holanda:2012ia} \\ \hline
\multirow{5}{*}{SNIa + SGL}  &  JLA \cite{SDSS:2014iwm} + SGL \cite{Cao:2015qja} & $-0.005^{+0.351}_{-0.215}$ &  && & \cite{Liao:2015uzb} \\ 
& \multirow{4}{*}{Union 2.1 \cite{Suzuki:2012apj} + SCP \cite{Rubin:2013apj} + SGL \cite{Cao:2015qja}} & $0.05 \pm 0.15$ (SIS, $\Lambda$)  & $0.09 \pm 0.30$ &  & & \multirow{4}{*}{\cite{Holanda:2015zpz}} \\ 
& & $0.08 \pm 0.22$ (PLaw, $\Lambda$) & $0.06 \pm 0.33$ & & &  \\
& & $0.01 \pm 0.22$ (SIS, $\omega$) & $0.017 \pm 0.28$ && &\\
& & $0.054 \pm 0.29$ (PLaw, $\omega$) & $0.0035 \pm 0.30$ && &\\ \hline 
\multirow{2}{*}{SNIa + BAO + CMB} & ESSENCE \cite{Davis:2007na} + SDSS+2dFGS \cite{Percival:2007yw}&  &  && $0.95 \pm 0.025$ & \multirow{2}{*}{\cite{Lazkoz:2007cc}} \\ 
&  + WMAP3 \cite{Wang:2006ts}  &  &  &&  ($1<z<2$)&  \\ \hline 
SNIa + BAO + GWs & LSST + ET simulation &  &  && $0.977 \pm 0.033$  & \cite{Hogg:2020ktc} \\ \hline
\end{tabular}
\end{table}

\section{Analysis and results}
\label{sec:Anal}

In order to improve the constraints, we use measurements from various probes. We analyze a set of the latest data providing information on angular diameter distances and the luminosity distance in order to constrain the deviation from the standard CCDDR. This is given by $-b/8$ in meVSL model as shown in Eq.~\eqref{etameVSL}. 

\subsection{SNIa data}
\label{subsec:SNIa}

We use the updated largest spectroscopically confirmed Pantheon compilation of 1048 SNIa data in 40 bins compressed for the luminosity distance in the redshift range $0.01 < z < 2.3$ \cite{Scolnic:2017caz}. We use the systematic covariance $\textbf{C}_{\text{SN}}$ for a vector of binned distances 
\begin{align}
\textbf{C}_{\text{SN} ij} = \sum_{n=1}^{N} \left( \frac{\partial \mu_{i}}{\partial S_{n}} \right)  \left( \frac{\partial \mu_{j}}{\partial S_{n}} \right) \sigma_{S_k} \label{CSN} \,, 
\end{align}
 where the summation is over the $n$ systematic with $S_n$ and its magnitude of its error $\sigma_{s_{n}}$. The $\chi^2$-analysis for Pantheon SNIa data is given by
 \begin{align}
 \chi^2_{\text{SN}} = \sum_{ij} \Delta \mu_{i}^{T} \cdot \left( C_{\text{SN}}^{-1} \right)^{ij} \cdot \Delta \mu_{j} \label{chiSN} \,,
 \end{align}
 where $\Delta \mu = \mu_{\text{data}} - \mu_{\text{obs}} - M$ with $M$ is a nuisance parameter. The covariance matrix includes both systematic and statistics with a diagonal component.   
From s10052-019-7263-9.pdf

\subsection{CMB data}
\label{subsec:CMB}

We also use the latest Planck data release \cite{Planck:2018vyg} in the condensed form of the so-called CMB shift parameters \cite{Wang:2007mza}
\begin{align}
R &\equiv \sqrt{\Omega_{m0} H_{0}^2} r(z_{\ast}) /\tc_0 \label{Rshift} \,, \\
l_{a} &\equiv \pi r(z_{\ast})/r_{s}(z_{\ast}) \label{lashift} \,, 
\end{align}
where $r(z_{\ast})$ is the comoving distane from the observer to the photon-decoupling epoch, $z_{\ast}$ and $r_{s}(z_{\ast})$ corresponds the comoving sound horizon at $z_{\ast}$. There is no modification to $R$ of meVSL model compared to that of GR. However, $l_{a}$ of meVSL is modified compared to that of GR due to the modification in the sound speed

\begin{align}
c_{s}^2 \equiv \left( \frac{\partial P}{\partial \rho} \right) =  \left( \frac{d P_{\gamma}}{d \rho_{\gamma}} \frac{\partial \rho_{\gamma}}{\partial \rho} \right) 
= \frac{1}{3} \tc_{0}^2 a^{b/2} \left( 1 + \frac{3+b/2}{4+b/2} \frac{\rho_{b}}{\rho_{\gamma}} \right)^{-1} \label{cs2meVSL} \,.
\end{align} 
This modification of $c_{s}$ causes the modification on the comoving sound horizon 
\begin{align}
r_{s}(z_{\ast}) = \frac{\tc_0}{H_0} \int_{z_{\ast}}^{\infty} \frac{1}{\sqrt{3}} \left( 1 + \frac{3+b/2}{4+b/2} \frac{\rho_{b}}{\rho_{\gamma}} \right)^{-1/2}\frac{dz}{E^{(GR)}(z)} \label{rsmeVSL} \,.
\end{align}
It has been known that these two shift parameters ({\it a.k.a.}, distance priors), $R$ and $l_a$, along with $\omega_{b} = \Omega_{b0} h^2$, provide an efficient summary of CMB data as far as dark energy constraints concern. The spectral scalar index $n_{s}$ is often included for completeness in addition to distance priors. 

We use CMB temperature and polarization data as well as lensing in the Planck data. Especially, we use the \sloppy{base$\_$plikHM$\_$TTTEEE$\_$lowl$\_$lowE$\_$lensing}  in the base MCMC chain to obtain the compressed Planck data for a spatially flat model. The final result are given by Gaussian distributions of a data vector $\textbf{v} = (l_{a}, R,  \omega_{b}, n_s)^{\text{T}}$ with the following means and standard deviations as well as its normalized covariance matrix 
\begin{align}
\textbf{v} = \begin{pmatrix} 301.80845 \\ 1.74963 \\  0.02237 \\ 0.96484 \end{pmatrix} \quad , \quad \sigma_{\text{v}} = \begin{pmatrix} 0.09007 \\ 0.00400 \\  0.00015 \\ 0.00415 \end{pmatrix} \quad , \quad\textbf{C}_{\text{v}} = \begin{pmatrix} 1.00000 & 0.47513 & -0.37693 & -0.31875 \\ 0.47513 & 1.00000 & -0.62305 & -0.67579 \\ -0.37693 & -0.62305 & 1.00000 & 0.39279 \\ -0.31875 & -0.67579 & 0.39279 & 1.00000 \end{pmatrix} \label{vandCv} \,.
\end{align}

When we marginalize the CMB distance priors over $n_s$, the covariance matrix for other three quantities $(l_{a}, R,  \omega_{b})$ becomes
\begin{align}
\textbf{C}_{\text{mar}} (p_{i},p_{j}) = \sigma(p_i) \sigma(p_j) \textbf{C}_{\text{v}-1} \label{Cmar}  \,,
\end{align}
where $j,j = 1,2,3$ and $\textbf{C}_{\text{v}-1}$ is the normalized covariance matrix dropping the 4th row and the 4th column from $\textbf{C}_{\text{v}}$. 

We consider a maximum likelihood determined by a $\chi^2$ statistics for CMB distance priors 
\begin{align}
\chi^2_{\text{CMB}} = \sum_{i=1}^{3} \sum_{j=1}^{3} \Delta p_{i} \left( \textbf{C}_{\text{mar}}^{-1} \right)^{ij} \Delta p_{j} \quad , \quad \Delta p_{i} = p_{i} - p_{i}^{\text{data}} \label{chiCMB} \,,
\end{align}
where $p_{i}^{\text{data}}$ are given in Eq.~\eqref{vandCv} and $\textbf{C}_{\text{mar}}^{-1}$ is the inverse matrix of the marginalized covariance matrix  given in Eq.~\eqref{Cmar}. 


 
\subsection{H data}
\label{subsec:H}

The expansion rate of the Universe is measured by the Hubble parameter, $H(z)$ . One can obtain it at the different epoch by using the cosmic chronometers approach from Baryon Oscillation Spectroscopic Survey (BOSS) Data \cite{Guo:2015gpa,Moresco:2016mzx}. We use 36 data points of the $\chi^2$ analysis for $H(z)$ 
\begin{align}
\chi^2_{H} (\textbf{p})=  \sum_{i=1}^{36} \frac{\left( H^{\text{th}}(z_i ; \textbf{p}) - H^{\text{obs}}(z_i) \right)^2}{\sigma_{H i}^2} \label{chiH} \,,
\end{align}
where $z_i$ is the redshift where $H$ has been measured, $\textbf{p}$ denote cosmological parameters, $H^{\text{th}(\text{obs})}(z_i)$ is the theoretical (observed) value of $H$ at $z_i$ and $\sigma_{H i}$ is the standard deviation of the $i$-th point.  

\subsection{Results}
\label{subsec:results}

We perform the maximum likelihood analysis for standard candles (SNIa) and standard rulers (CMB) along with Hubble parameters assuming flatness to put the constraints on cosmological parameters. We investigate separately SNIa, CMB distance priors, and H data and analyze minimization of the corresponding $\chi^2$ with respect to cosmological parameters $h, \Omega_{m0}, \omega$, and $b$. We also perform $\chi^2$ for $h$ and $b$ with various priors of $\Omega_{m0}$ for $\Lambda$CDM models. We summarize these results in Table.~\ref{tab:my-table2}. $\bar{\chi}^2$ denotes the reduced $\chi^2$. First, we perform the joint $\chi^2$-analysis for both CMB data with Hubble parameters data. After we perform $\chi^2$ analysis for four parameters ($h, \Omega_{m0}, \omega, b$), we marginalize with respect to the best fit value of $\omega$ to obtain the 1-$\sigma$ confidence level values of other parameters. The best fit value of $\omega = -1.0715$ and we obtain following means and standard deviations for other parameters
\begin{align}
\langle h \rangle = 0.6971 \quad &, \quad \sigma_{h} = 0.0041 \nonumber \,, \\
 \langle \Omega_{m0} \rangle = 0.2968 \quad &, \quad \sigma_{\Omega_{mo}} = 0.0055 \label{CMBHb} \,, \\ 
 \langle b \rangle = 0.1410 \quad &, \quad \sigma_{b} = 0.0869 \nonumber \,, 
\end{align}
with the reduced $\chi^2$ as 0.557. This might be over-fitted. We perform $\chi^2$-analysis for the joint data with $H$ and the Pantheon. The result is shown in the second row of Table~\ref{tab:my-table2}. The best fit value of $\omega = -1.018$ and means and standard deviations for others are obtained as
\begin{align}
\langle h \rangle = 0.7053 \quad &, \quad \sigma_{h} = 0.0219 \nonumber \,, \\
 \langle \Omega_{m0} \rangle = 0.2839 \quad &, \quad \sigma_{\Omega_{mo}} = 0.0180 \label{HSNIab} \,, \\ 
 \langle b \rangle = 0.1577 \quad &, \quad \sigma_{b} = 0.1129 \nonumber \,, 
\end{align}
with the reduced chi-squared value is 0.976. This indicates that the extent of the match between estimations and observations is in good accord with the error variance. 

We also perform $\chi^2$ analysis using CMB, H, and, SNIa data. In this case, the best-fit value of $\omega$ is consistent with $\Lambda$CDM. Thus, we limit our analysis to $\omega = -1$ and put priors on other parameters in order to investigate constraints on specific parameters. For $\Omega_{m0} = 0.274$, we obtain mean values and marginalized $68$\% confidence level errors for $h$ and $b$ are given by 
\begin{align}
\langle h \rangle = 0.7059 \quad &, \quad \sigma_{h} = 0.0006 \nonumber \,, \\
 \langle b \rangle = 0.1353 \quad &, \quad \sigma_{b} = 0.0760 \label{CMBHSNIaOm27b} \,, 
\end{align}
with $\bar{\chi}^2 = 1.032$. If we put $\Omega_{m0} = 0.314$, then the obtained mean and standard deviations are
\begin{align}
\langle h \rangle = 0.6766 \quad &, \quad \sigma_{h} = 0.0006 \nonumber \,, \\
 \langle b \rangle = 0.0896 \quad &, \quad \sigma_{b} = 0.0740 \label{CMBHSNIaOm31b} \,, 
\end{align}
with $\bar{\chi}^2 = 0.976$. This analysis shows the 1-$\sigma$ CL deviation of CDDR. Without any prior on $\Omega_{m 0}$, we obtain
\begin{align}
\langle h \rangle = 0.6761 \quad &, \quad \sigma_{h} = 0.0036 \nonumber \,, \\
 \langle \Omega_{m0} \rangle = 0.3141\quad &, \quad \sigma_{\Omega_{mo}} = 0.0054 \label{CMBHSNIab} \,, \\ 
 \langle b \rangle = 0.0010 \quad &, \quad \sigma_{b} = 0.0724 \nonumber \,, 
\end{align}
with $\bar{\chi}^2 = 0.978$. Compared to other case, this result shows no deviation of CDDR within 1-$\sigma$ CL. 

\begin{table}[]
\caption{}
\label{tab:my-table2}
\begin{tabular}{|c|c|c|c|c|c|}
\hline
 & $h$ & $\Omega_{m0}$ & $\omega_0$ & $b$ & $\bar{\chi}^2$ \\ 
 \hline
CMB + H & $0.6971 \pm 0.0041$ & $0.2968 \pm 0.0055$ & $-1.0715$ & $0.1410 \pm 0.0869$ & 0.557 \\
H + SNIa & $0.7053 \pm 0.0219$ & $0.2839 \pm 0.0180$ & $-1.0180$  & $0.1577 \pm 0.1129$  & 0.976 \\ 
 \multirow{3}{*}{CMB + H + SNIa} & $0.7059 \pm 0.0006$& $0.274$ & $-1$ & $0.1353\pm 0.0760$ & $1.032$ \\ 
 & $0.6766 \pm 0.0006$& $0.314$ & $-1 $ & $0.0896 \pm 0.0740$ & $0.976$ \\
 & $0.6761 \pm 0.0036$& $0.3141 \pm 0.0054 $ & $-1 $ & $0.0010 \pm 0.0724$ & $0.978$ \\  
\hline
\end{tabular}
\end{table}

\begin{figure*}
\centering
\vspace{1cm}
\begin{tabular}{cc}
\includegraphics[width=0.5\linewidth]{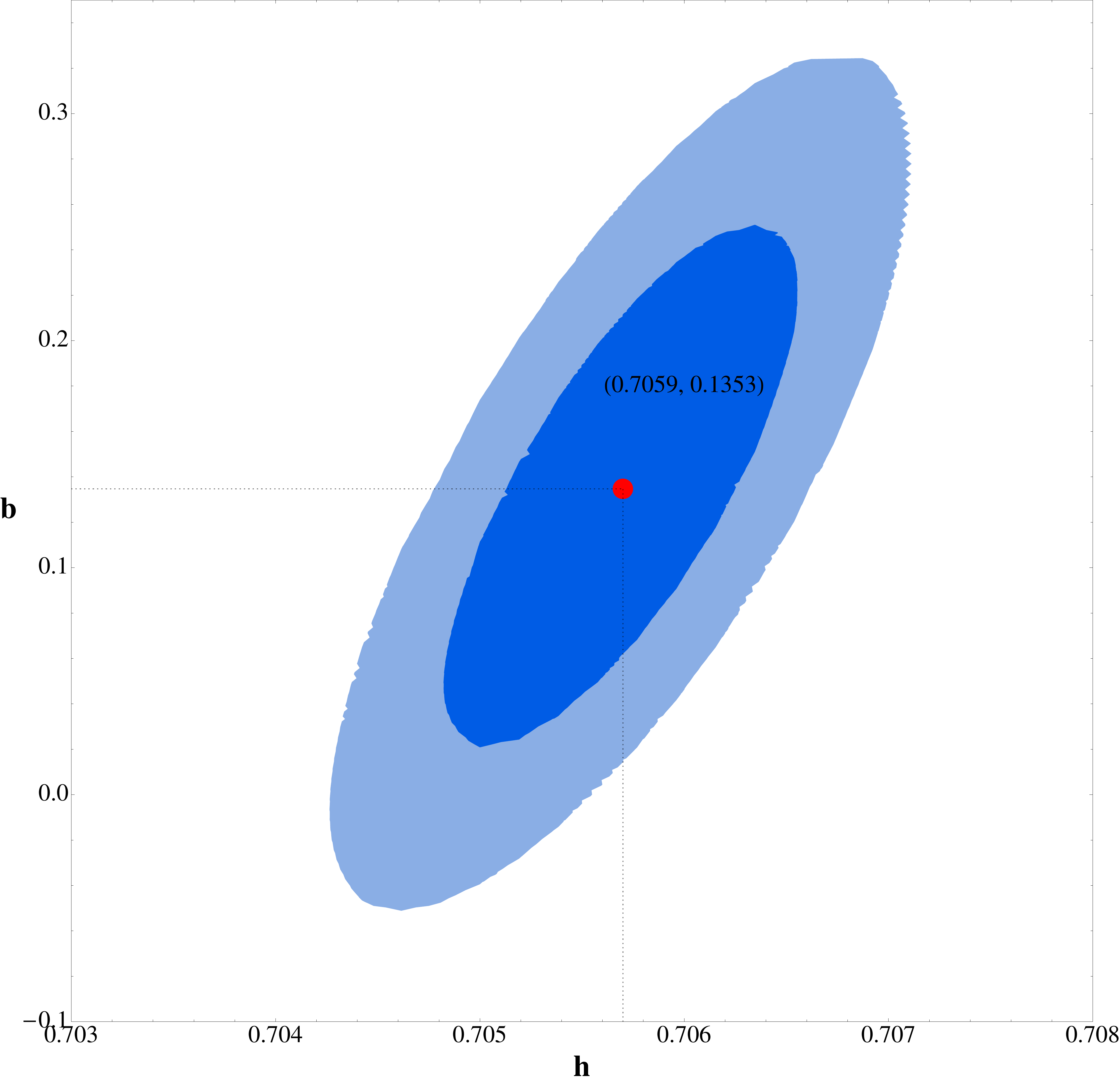} &
\includegraphics[width=0.49\linewidth]{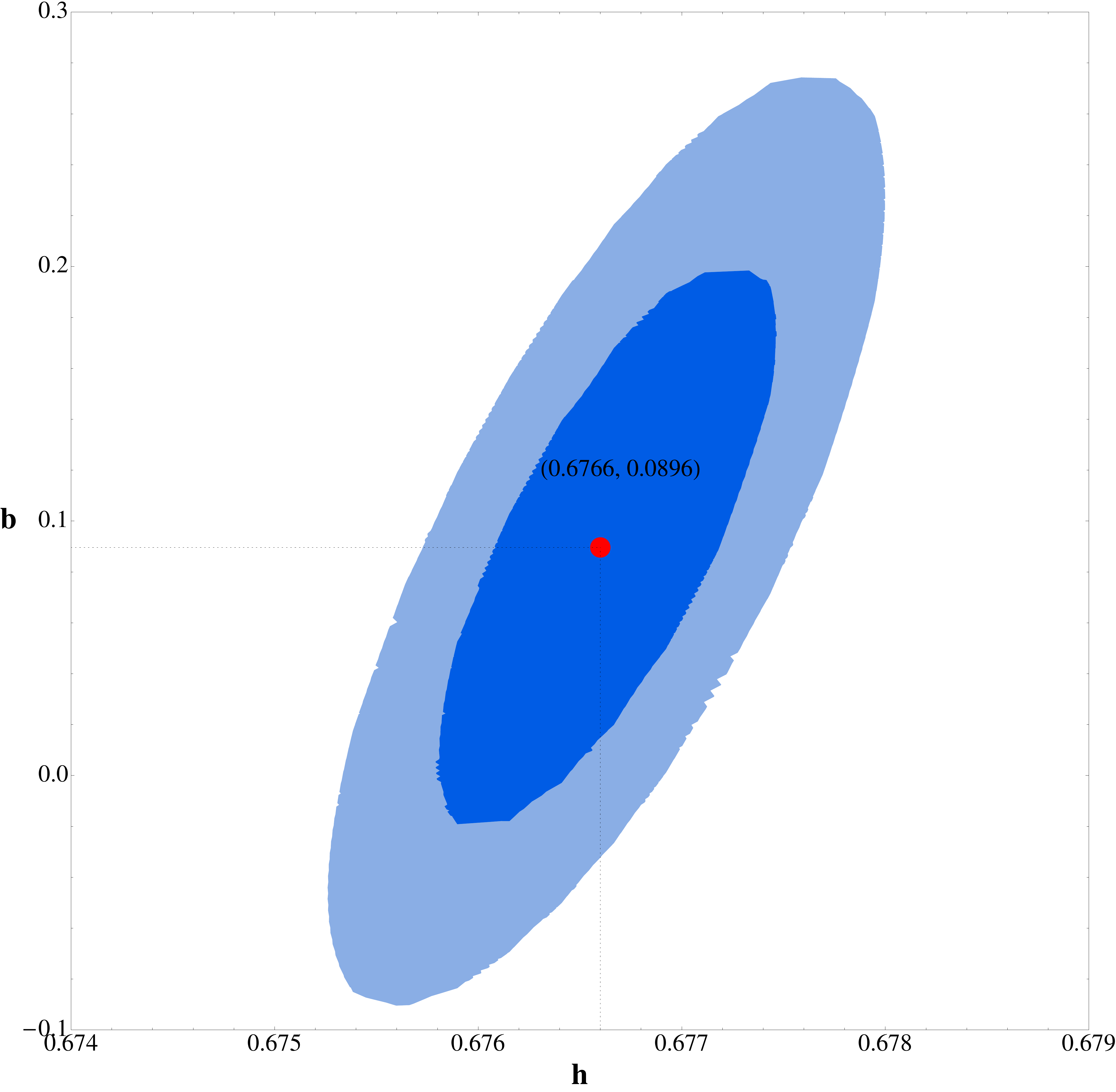}
\end{tabular}
\vspace{-0.5cm}
\caption{The joint 68 \% and 95 \% confidence contour for ($h, b$) from CMB, H, and SNIa with prior on the value of $\Omega_{m0}$. a) With $\Omega_{m0} = 0.274$. b) $\Omega_{m0} = 0.314$. } \label{fig-1}
\vspace{1cm}
\end{figure*}

We also show this result in Figure~\ref{fig-1}. Fig.~\ref{fig-1} shows the joint 1-$\sigma$ and 2$\sigma$ confidence contours for ($h, b$) from Planck, H chronometers, and Pantheon data. The darker and lighter regions represent 1-$\sigma$ and 2-$\sigma$ CL regions, respectively. The left panel of this figure corresponds to the case with the prior on the value of $\Omega_{m0}$ as $0.274$. Within $68$ \% CL, the value of $b$ is above zero to indicate the deviation from CDDR. When we fix $\Omega_{m0} = 0.314$, the $68$ \% and $95$ \% confidence contours for ($h, b$) are shown in the right panel of Fig.~\ref{fig-1}. However, the current data is consistent with the standard model in this case even within $1$-$\sigma$ CL.

\section{Conclusions}
\label{sec:Con}

We have constrained deviations from the standard distance duality relation (CDDR) due to the minimally extended varying speed of light model using current data. We have used the latest supernovae and cosmic microwave background radiation data with $38$ Hubble parameters data obtained from cosmic chronometers. In addition to the previously known mechanism for the violation of CDDR, we also introduce another mechanism due to the varying speed of light in this manuscript. Especially, the deviation relation in this model is consistent with one of the previously used parameterizations.  

We also have derived the CMB distance priors from the last Planck data. We obtain mean values and the normalized covariance matrix of {$l_a, R, \omega_b, n_s$} by using CMB temperature and polarization data with lensing.

We have performed the maximum likelihood analysis for joint data. Some of them indicate that current data show the 1-$\sigma$ deviation from the standard CDDR. However, with different priors on some cosmological parameters, the current data is consistent with the standard model without showing any deviation from the cosmic duality distance relation. Thus, we may need more accurate data to investigate any deviation of CDDR. We also conclude that meVSL is a still viable model and we have obtained another constraint on the value of $b$ from CDDR in addition to previous constraints on it from other probes \cite{Lee:2021ona,Lee:2021jcg}.

\section*{Acknowledgments}
SL is supported by Basic Science Research Program through the National Research Foundation of Korea (NRF) funded by the Ministry of Science, ICT, and Future Planning (Grant No. NRF-2017R1A2B4011168).


\end{document}